\documentclass[sort&compress
    ,final            
  ,4apaper]
  {aipproc}
\usepackage{caption}
\usepackage{bm}
\usepackage{color}
\usepackage{amsmath}
\usepackage{amssymb}
\usepackage{graphicx}

    \usepackage{showkeys}

\layoutstyle{8x11single}

\newcommand\beq{\begin{equation}}
\newcommand\eeq{\end{equation}}
\newcommand\beqa{\begin{eqnarray}}
\newcommand\eeqa{\end{eqnarray}}

\begin{document}

\title{Properties of the Homogeneous Cooling State of a Gas of Inelastic Rough Particles}


\author{Francisco Vega Reyes}{
  address={Departamento de F\'isica, Universidad de Extremadura, E-06071 Badajoz, Spain},
altaddress={Instituto 
de Computación Científica Avanzada 
(ICCAEx), Universidad de Extremadura, E-06071 Badajoz, Spain}
}

\author{Andr\'es Santos}{
  address={Departamento de F\'isica, Universidad de Extremadura, E-06071 Badajoz, Spain},
altaddress={Instituto 
de Computación Científica Avanzada 
(ICCAEx), Universidad de Extremadura, E-06071 Badajoz, Spain}
}

\author{Gilberto M. Kremer}{
  address={Departamento de F\'isica, Universidade Federal do Paran\'a, Curitiba, Brazil}
}

\begin{abstract}
 {In this work we address the question of whether}  a  {low-density} system composed of  identical  {rough} particles may reach hydrodynamic states  (also called \textit{normal} states),  {even if energy is not conserved  in particle collisions}. As a way to measure the ability of the system to    {present a hydrodynamic behavior}, we  focus on the so-called homogeneous cooling state of the granular gas and look at the corresponding relaxation time as a function of inelasticity and roughness. We report computer simulation results of the  sixth- and eighth-order cumulants of the particle  {velocity} distribution function  {and study the influence of roughness on their relaxation times and  asymptotic values}.  This extends the results of a previous work  {[Phys. Rev. E \textbf{89}, 020202(R) (2014],} where   {lower-order} cumulants were measured. Our results  confirm that  {the} relaxation times are not necessarily longer for stronger inelasticities. This implies that inelasticity by itself does not preclude hydrodynamics.
 It is also observed that the cumulants associated with the angular velocity distribution may reach very high values in a certain region of (small) roughness and that these maxima coincide with small orientational correlation points.
\end{abstract}

\maketitle


\section{Introduction}
\label{sec1}

Matter is inherently a many-body system. Logically, for a system with a very large number of particles the accurate description of each entity's dynamics becomes irrelevant, whereas a statistical picture of the system can yield a meaningful {account} of the underlying physics \cite{P00}. In matter systems where particles are atoms or molecules, and if the internal dynamics is not relevant in the process of  {dynamical} exchange between particles (henceforth, \textit{collisions}), we may have two separated dynamics levels: a quantum one for atom dynamics description {and a classical one} for particle dynamics exchange description \cite{P00}. Therefore, under certain conditions, a classical {statistical-mechanical} theory is appropriate for the physics resulting from particles collisions. In this case, and for  {low-density fluids}, the classical Boltzmann equation applies \cite{CC70}.

Furthermore, the statistical description resulting from the Boltzmann equation is amenable to a higher level description called \textit{hydrodynamics} \cite{B53,B67,CL00}. In the hydrodynamic approach, low density states are characterized by a single particle distribution function (``normal'' solution) whose spatial dependence may be written entirely through  {a functional}  dependence on the relevant average fields. These fields appear in the appropriate balance equations of the system \cite{CL00,G03}. The three basic fields are density, flow velocity, and temperature, resulting from mass, momentum, and energy balance, respectively.

Let us now suppose that our system has indeed a large number of particles at low density but that these are not  atoms {or} molecules but much larger particles (typical size of at least $1~\mu$m), that we can call \textit{grains} \cite{B54}. Would then the hydrodynamic description work at a theoretical level? Would the behavior of this kind of systems look like   {that} of  {a classical} fluid? \cite{DB11}  {Obviously,} ``fluid'' properties are present in some peculiar way.  In fact, many phenomena present in classical fluids, such as instabilities, convection, diffusion, thermal segregation, laminar flow, etc., are analogously present in low-density \textit{granular} systems  \cite{G03,JNB96}. Thus, at a theoretical and fundamental level  an important question  {arises}: are the equations of fluid mechanics appropriate for a many-particle system at low density, independently of the size of its particles, even if collisions dissipate energy? At a ``real life'' level the answer to this question is  important and far reaching since a positive answer would imply that we can find fluid-like behavior  {(and apply the appropriate tools)} in a wealth of very common systems (sands, asteroid rings, powders, stone avalanches, cereal mills, etc.) \cite{JNB96}.

The aim of this work is to support the growing evidence  that, indeed, fluid mechanics should in principle be appropriate for low-density granular systems without \textit{a priori} limitation on the degree of inelasticity in particle collisions \cite{DB11},  {even if particle roughness is significant}. 
Our results will also allow us to show that,  {on the other hand,} the distribution function of the \textit{granular gas} of rough spheres may strongly deviate from the Maxwellian.

 {In} this work, we consider a system with a large number of identical particles  {that} may eventually collide loosing a fraction of their kinetic energy. The system is at very low density and thus collisions are always binary and instantaneous. Therefore, our system is a \textit{granular gas}  {that} may be described by the inelastic versions of the Boltzmann and/or Enskog equations \cite{BP04,GDH07,VSG10,VSK14}. We let our particle velocities have translational and rotational degrees of freedom. For simplicity, we will assume that the momentum transfer in collisions occurs with no sliding between the particles. Therefore, collisions may be characterized by  {the coefficients of normal and tangential restitution   \cite{BPKZ07}, measuring the degrees of inelasticity and roughness, respectively}. We also assume that those coefficients  are constant (i.e., they do not depend on the  velocities of the colliding particles).

For this rough hard-sphere model, we show by computer simulations that, after a transient, the {\emph{undriven}} system relaxes always to a ``normal'' (or hydrodynamic) state, the so-called homogeneous cooling state (HCS), the relaxation stage length being practically not related to the degree of inelasticity (coefficient of normal restitution) in the collisions. Furthermore, in the range of interest for experiments  {(medium or large roughness)}, this relaxation time does not present a clear tendency with respect to the degree of roughness of the particles {(coefficient of tangential restitution)} either. However, the system seems to have very long relaxation times near the smooth collision limit ({i.e., at small} particle roughness).
 {Moreover, strong non-Maxwellian effects are present within a certain range of roughness. These findings are based on measurements of sixth- and eighth-order velocity cumulants, thus complementing a previous study based on the fourth-order cumulants \cite{VSK14}.}

\section{The inelastic rough hard-sphere model}
\label{sec2}
We consider  {in a volume $V$} a set of  {$N=nV$} identical  hard spheres with diameter $\sigma$, mass $m$,  {and} moment of inertia $I$.  {Each sphere possesses a translational and an angular velocity} $\mathbf{v}$ and $\bm{\omega}$, respectively,  {and collisions are binary and instantaneous}.
In addition, the spheres are dissipative, i.e., a fraction of kinetic energy is lost during collisions.

The collisional model we will use is the inelastic rough hard-sphere model \cite{CC70,BPKZ07,GNB05},  {in which} collisions are characterized by two (constant) coefficients of restitution: $\alpha$ for normal restitution  {and $\beta$} for tangential restitution \cite{VSK14}:
\begin{equation}
\widehat{\bm{\sigma}}\cdot\mathbf{v}'_{12}=-\alpha\widehat{\bm{\sigma}}\cdot\mathbf{v}_{12}, \quad  \widehat{\bm{\sigma}}\times\mathbf{v}'_{12}=-\beta\widehat{\bm{\sigma}}\times\mathbf{v}_{12},
\end{equation}
where  {$\widehat{\boldsymbol{\sigma}}$ is a unit vector
directed along the centers of the two colliding particles,} $\mathbf{v}_{12}$ is the relative velocity of the
contact points of the colliding pair of particles, and primes denote post-collisional values. The  coefficient  {of normal restitution} ranges from $\alpha=0$ (completely inelastic  {collision}) to $\alpha=1$ (perfectly elastic  {collision}), while the  coefficient of  {tangential} restitution  {(or ``roughness'' parameter)} ranges from $\beta=-1$ ({perfectly smooth particles}, with no reversal of tangential relative velocity) to $\beta=1$ ({perfectly rough particles}, with perfect reversal of tangential relative velocity).  For a complete description of the collisional rules,  {we  refer the reader to Refs.\ \cite{BPKZ07,SKS11}}.

Under  {spatially homogeneous} conditions, the Enskog kinetic equation in its inelastic version applies,
\begin{equation}
\frac{\partial f}{\partial t}= {\chi(n)J}[\mathbf{v},\bm{\omega}|f], \label{KE}
\end{equation}
where  {$f(\mathbf{v},\bm{\omega},t)$ is the one-body velocity distribution function,} $\chi(n)$ is the pair correlation function at contact (Enskog factor) \cite{HM06}, and $J$ is the  {Boltzmann} collision operator for inelastic rough hard spheres (for an explicit expression of this operator, see Ref.\ \cite{SKS11}).
 {Given a certain dynamical variable $A(\mathbf{v},\bm{\omega})$, its average value is given by}
\beq
 {\langle A\rangle=\frac{1}{n}\int d\mathbf{v}\int d\bm{\omega}\, A(\mathbf{v},\bm{\omega})f(\mathbf{v},\bm{\omega},t)}.
\eeq
 {As usual \cite{BPKZ07,SKS11}, the translational and rotational temperatures are defined  as}
\beq
 {T_t=\frac{m}{3}\langle(\mathbf{v}-\mathbf{u}_f)^2\rangle,\quad T_r=\frac{I}{3}\langle\omega^2\rangle,}
\label{TrTt}
\eeq
 {respectively,
where  $\mathbf{u}_f=\langle \mathbf{v}\rangle$ is the flow velocity. Due to energy dissipation, the total temperature $T=\frac{1}{2}(T_t+T_r)$ decays monotonically with time (unless $\alpha=\beta^2=1$). Since for the kinetic equation \eqref{KE} both the density $n$ and $\mathbf{u}_f$ are constant, $T$ is the only relevant hydrodynamic quantity of the system. Thus, if a hydrodynamic regime {does} exist, the whole temporal dependence of $f$ {must} occur through a dependence on $T$ \cite{vNE98}.}

 {It is then convenient to rescale the translational and angular velocities with the respective temperatures and define the reduced quantities}
\beq
 {\mathbf{c}(t)\equiv\frac{\mathbf{v}-\mathbf{u}_f}{\sqrt{2T_t(t)/m}}}, \quad
 {\mathbf{w}(t)\equiv\frac{\bm{\omega}}{\sqrt{2T_r(t)/I}}}.
\label{cw}
\eeq
 {Consequently, the reduced distribution function is}
\beq
 {\phi(\mathbf{c},\mathbf{w},t)\equiv \frac{1}{n}\left[\frac{4T_t(t)T_r(t)}{m I}\right]^{3/2}f(\mathbf{v},\bm{\omega},t)}.
 \eeq
 {In this undriven and spatially homogeneous situation, the existence of a hydrodynamic state (the HCS) implies that, at given $\alpha$ and $\beta$, both the temperature ratio $\theta(t)\equiv T_r(t)/T_t(t)$ and the reduced distribution function $\phi(\mathbf{c},\mathbf{w},t)$ reach well defined stationary values after a certain relaxation time. This expectation was recently confirmed \cite{VSK14} at the level of $\theta(t)$ and of the fourth-order velocity cumulants by computer simulations and a Grad approximation. As said in the Introduction, our main aim here is to extend the simulation test to sixth- and eighth-order cumulants, checking whether the relaxation times are sensitive or not to the order of the cumulants.}

\section{Velocity cumulants}
\label{sec3}
 {Now we assume that the state is \emph{isotropic}. This means that $\phi(\mathbf{c},\mathbf{w},t)$ must be a function of the three scalar quantities $c^2=\mathbf{c}\cdot\mathbf{c}$, $w^2=\mathbf{w}\cdot\mathbf{w}$, and $(\mathbf{c}\cdot \mathbf{w})^2$, so we can represent it as a polynomial expansion of the form \cite{VSK14,SK12}}
\beq
\phi(\mathbf{c},\mathbf{w},t)= {\frac{e^{-c^2-w^2}}{\pi^{3}}}\sum_{j=0}^\infty\sum_{k=0}^\infty\sum_{\ell=0}^\infty a_{jk}^{(\ell)}(t)\Psi_{jk}^{(\ell)}(\mathbf{c},\mathbf{w}),
\label{phi}
\eeq
where
\beq
\Psi_{jk}^{(\ell)}(\mathbf{c},\mathbf{w})=L_j^{(2\ell+\frac{1}{2})}(c^2) L_k^{(2\ell+\frac{1}{2})}(w^2)\left(c^2w^2\right)^\ell
P_{2\ell}(u)
\eeq
is a polynomial of total degree   $2(j+k+2\ell)$ in velocity.
Here, $L_j^{(\alpha)}(x)$ and $P_{2\ell}(x)$ are Laguerre  and Legendre polynomials, respectively, and
$u\equiv (\mathbf{c}\cdot \mathbf{w})/cw$.  {The} set of polynomials $\{\Psi_{jk}^{(\ell)}\}$ constitute a complete orthogonal basis \cite{VSK14},  {i.e.,}
\beq
 {\langle \Psi_{jk}^{(\ell)}|\Psi_{j'k'}^{(\ell')}\rangle=N_{jk}^{(\ell)}\delta_{j,j'}\delta_{k,k'}\delta_{\ell,\ell'}},
\label{avPsi}
\eeq
 {where the normalization constants are }
\beq
 {N_{jk}^{(\ell)}= \frac{\Gamma(j+2\ell+\frac{3}{2})\Gamma(k+2\ell+\frac{3}{2})}{[\Gamma(\frac{3}{2})]^2(4\ell+1)j!k!},}
\eeq
 {and the scalar product of two  arbitrary isotropic (real) functions $\Phi(\mathbf{c},\mathbf{w})$ and $\Phi'(\mathbf{c},\mathbf{w})$ is defined as}
\beq
 {\langle \Phi|\Phi'\rangle\equiv\pi^{-3} \int d\mathbf{c}\int d\mathbf{w} \,e^{-c^2-w^2} \Phi(\mathbf{c},\mathbf{w})\Phi'(\mathbf{c},\mathbf{w})}.
\eeq

{Because of the normalization condition of $\phi$, $a_{00}^{(0)}=1$. Also, the definitions \eqref{TrTt} and \eqref{cw} imply that $\langle c^2\rangle=\langle w^2\rangle=\frac{3}{2}$, so that $a_{10}^{(0)}=a_{01}^{(0)}=0$. The rest of the expansion coefficients $a_{jk}^{(\ell)}$ (with $j+k+2\ell\geq 2$) represent the velocity cumulants of order $2(j+k+2\ell)$ and measure the deviation of the reduced distribution function from the Maxwellian. Thanks to the property \eqref{avPsi}, the cumulants are given by}
\beq
 {a_{jk}^{(\ell)}=\frac{\langle\Psi_{jk}^{(\ell)}\rangle}{N_{jk}^{(\ell)}}} .
\eeq
 {Therefore, $a_{jk}^{(\ell)}$ is a linear combination of the moments $\langle c^{2p}w^{2q}
(\mathbf{c}\cdot\mathbf{w})^{2r}\rangle$ with  $p+q+2r\leq j+k+2\ell$.} In particular, the fourth-order ($j+k+2\ell=2$) cumulants are
\begin{equation}
a_{20}^{(0)}=\frac{4}{15}\langle c^4\rangle-1, \quad a_{02}^{(0)}=\frac{4}{15}\langle w^4\rangle-1,
 \eeq
 \beq
a_{11}^{(0)}=\frac{4}{9}\langle c^2w^2\rangle-1,\quad a_{00}^{(1)}=\frac{8}{15}\left[\langle (\mathbf{c}\cdot\mathbf{w})^2\rangle-\frac{1}{3}\langle c^2w^2\rangle\right].
\label{a20}
\end{equation}
 {In Ref.\ \cite{VSK14} we constructed a Grad theory by truncating the expansion \eqref{phi} after the fourth-order cumulants and neglecting nonlinear terms in the collision operator.}

If a hydrodynamic state is to be reached, then the scaled distribution function {$\phi$} becomes independent of time since all the  {associated} cumulants are also stationary (all of them are reduced with thermal velocities). Thus, we measured in our previous work \cite{VSK14}  {(by Grad theory and by computer simulations)} the relaxation times (duration of the transient before the steady state is reached) of the fourth-order cumulants and of the temperature ratio $\theta$. The results we obtained showed that the hydrodynamic solution relaxation time is  {practically insensitive to} inelasticity  {(i.e., the value of the coefficient $\alpha$). However, the relaxation time takes increasingly higher (finite) values as one approaches the nearly-smooth limit ($\beta\to -1$)}.
Additionally, we showed that in  {a ``critical''} region  {of small roughness (but not in the limit $\beta\to -1$)} the fourth-order cumulants (most notably, $a_{02}^{(0)}$) may become very large, signaling that the distribution function strongly  {deviates} in this region from the Maxwellian. On the other hand, even in this region, the relaxation times are finite and thus the hydrodynamic solution is always reached.

A natural question is whether the above conclusions remain being valid when higher-order cumulants are considered. For this reason, we have used the direct simulation Monte Carlo (DSMC) method \cite{B94} to measure all the sixth- and eighth-order cumulants with $\ell=0$, i.e., $a_{jk}\equiv a_{jk}^{(0)}$ with $j+k=3$ and $j+k=4$. More specifically,
the sixth-order cumulants are defined by
\begin{equation}
  a_{30}=-\frac{8}{105}\langle c^6\rangle +\frac{4}{5}\langle c^4\rangle -2, \quad a_{03}=-\frac{8}{105}\langle w^6\rangle+\frac{4}{5}\langle w^4\rangle -2, \label{a30}
\end{equation}
\begin{equation}
a_{21}= -\frac{8}{45}\langle c^4w^2\rangle +\frac{4}{15}\langle c^4\rangle  {+\frac{8}{9}\langle c^2w^2\rangle}-2, \quad
a_{12}= -\frac{8}{45}\langle c^2w^4\rangle +\frac{4}{15}\langle w^4\rangle  {+\frac{8}{9}\langle c^2w^2\rangle}-2.
\end{equation}
The eighth-order cumulants are
\begin{equation}
a_{40}= \frac{16}{945}\langle c^8\rangle -\frac{32}{105}\langle c^6\rangle+\frac{8}{5}\langle c^4\rangle-3, \quad
a_{04}= \frac{16}{945}\langle w^8\rangle -\frac{32}{105}\langle w^6\rangle+\frac{8}{5}\langle w^4\rangle-3,
\end{equation}
\begin{equation}
a_{31}=
\frac{16}{315}\langle c^6w^2\rangle -\frac{8}{105}\langle c^6\rangle-\frac{8}{15}\langle c^4w^2\rangle+\frac{4}{5}\langle c^4\rangle  +\frac{4}{3}\langle c^2w^2\rangle-3,
\end{equation}
\begin{equation}
a_{13}=
\frac{16}{315}\langle c^2w^6\rangle -\frac{8}{105}\langle w^6\rangle-\frac{8}{15}\langle c^2w^4\rangle+\frac{4}{5}\langle w^4\rangle  +\frac{4}{3}\langle c^2w^2\rangle-3,
\end{equation}
\begin{equation}
a_{22}=
\frac{16}{225}\langle c^4w^4\rangle -\frac{16}{45}\langle c^4w^2\rangle -\frac{16}{45}\langle c^2w^4\rangle+\frac{4}{15}\langle c^4\rangle +\frac{4}{15} {\langle w^4\rangle} +\frac{16}{9}\langle c^2w^2\rangle-3.
\label{a22}
\end{equation}

\section{Results}

\subsection{Relaxation toward the HCS}

\begin{figure}
\begin{tabular}{cc}
 \includegraphics[height=0.43\textheight]{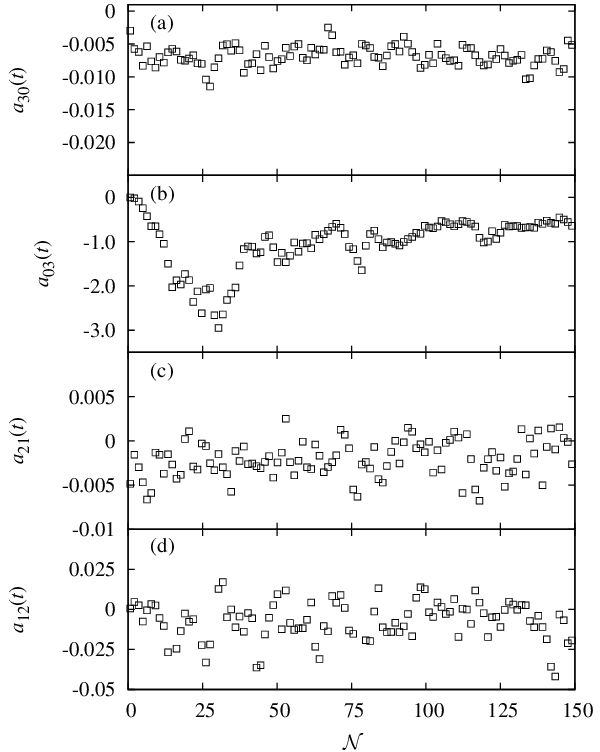}   & \includegraphics[height=0.43\textheight]{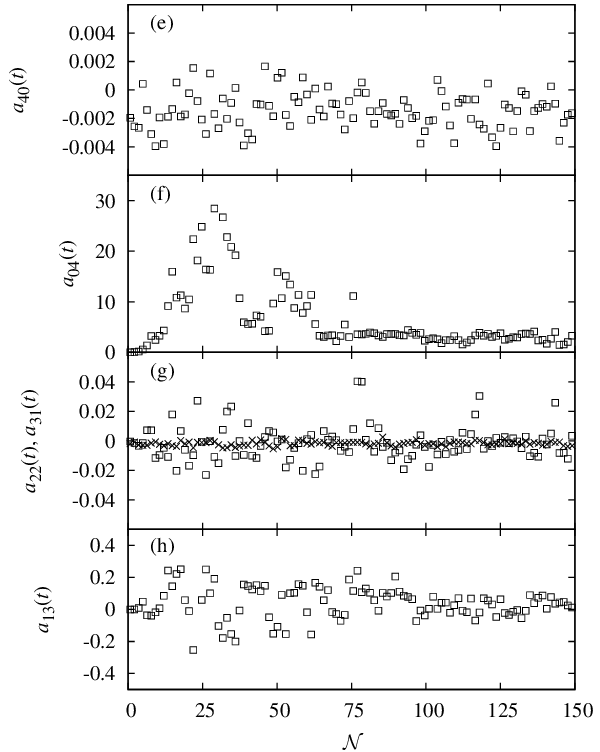}\\
    \caption{DSMC results for the  {time evolution of the sixth-order cumulants [panels (a)--(d)] and the  eighth-order cumulants [panels (e)--(h)] for a granular gas with $\alpha=0.9$, $\beta=-0.575$, and $I=m\sigma^2/10$}. Time is represented in units of the  {accumulated number of collisions per particle} ($\mathcal{N}$).   In panel (g), crosses and squares stand for $a_{31}$ and $a_{22}$, respectively. }
    \end{tabular}
 \label{fig1}
\end{figure}

We plot in Fig.\ \ref{fig1} the time evolution of the cumulants \eqref{a30}--\eqref{a22} for the inelastic rough hard-sphere granular gas in a representative case with  {$\alpha=0.9$, $\beta=-0.575$, and $I=m\sigma^2/10$} (particles with uniform mass density).  {The initial state was that of a Maxwellian distribution with equal temperatures, i.e., $\theta(0)=1$.}

As we can see,  {most cumulants are relatively small and, therefore, exhibit strong fluctuations.  On the other hand, the magnitudes of those cumulants with higher order in the rotational part (namely, $a_{03}$, $a_{13}$, and, especially, $a_{04}$) reach high values at about $\mathcal{N}\approx 30$ collisions per particle and then relax to their steady-state values with a relaxation time of about $\mathcal{N}_r\approx 100$ collisions per particle.}
These  {findings} are in accordance with previous results for the fourth-order cumulants \cite{VSK14}, where almost systematically the  cumulant $a_{02}$ was the one with the  {largest magnitude}.

{The temperature ratio and each} cumulant  have their own independent relaxation time,  conveniently defined as the value of $\mathcal{N}$ after which the deviations from the stationary value are smaller than 5\% \cite{VSK14}. We then define the \emph{global} relaxation time $\mathcal{N}_r$ as the maximum value obtained from the  set of  {relaxation times} measured for the cumulants and the temperature ratio. Let us call  {$\mathcal{N}_r^\text{I}$} to the  {value of $\mathcal{N}_r$ obtained from the temperature ratio and the fourth-order cumulants, as measured in our previous work \cite{VSK14}}. Similarly, we call   {$\mathcal{N}_r^\text{II}$} to the  {value of $\mathcal{N}_r$ obtained from} the set of sixth- and eighth-order cumulants. The  {interesting} question  {is whether}  {$\mathcal{N}_r^\text{II}$} is higher than  {$\mathcal{N}_r^\text{I}$} or, even more,  {whether $\mathcal{N}_r^\text{II}$} does not present finite values.

\begin{figure}
  \includegraphics[height=.25\textheight]{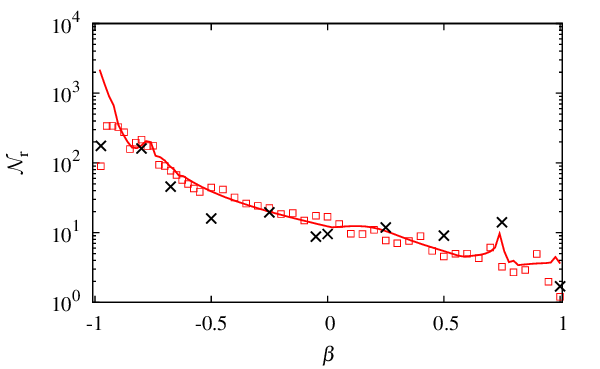}
  \caption{DSMC results for  {the relaxation times $\mathcal{N}_r^\text{I}$ and $\mathcal{N}_r^\text{II}$}, as functions of  {the roughness parameter} $\beta$,  for a granular gas with  {$\alpha=0.9$ and $I=m\sigma^2/10$}. {The red line (Grad theory) and the squares (DSMC) stand for $\mathcal{N}_r^\text{I}$, as obtained in Ref.\ \cite{VSK14} from the temperature ratio and the fourth-order cumulants}. Crosses stand for   {$\mathcal{N}_r^\text{II}$},  as extracted {in the present work} from the set of sixth- and eighth-order cumulants.}
  \label{fig2}
\end{figure}

{The two quantities $\mathcal{N}_r^\text{I}$ and $\mathcal{N}_r^\text{II}$ are compared in Fig.\ \ref{fig2} for $\alpha=0.9$ and $-1\leq\beta\leq 1$.} As can be seen, the results clearly show that, for the whole range of roughness $\beta$,  {$\mathcal{N}_r^\text{II}$} is not significantly  {different from $\mathcal{N}_r^\text{I}$}. This implies that the global description of the relaxation to hydrodynamics in our previous work remains  {essentially} the same. Thus, we can keep the conclusion that a hydrodynamic  {HCS} is always reached for a granular gas  {of rough particles. It was also observed in Ref.\ \cite{VSK14} that the values of $\mathcal{N}_r$ were rather insensitive to the degree of inelasticity (i.e., to the value of $\alpha$). Although we have not analyzed in Fig.\ \ref{fig2} a value different from $\alpha=0.9$, the good agreement found in that case reinforces the latter observation. Figure \ref{fig2} additionally shows that the relaxation time is relatively short ($\mathcal{N}_r\lesssim 10$) for the typical roughness values ($\beta\gtrsim 0$) of experimental relevance \cite{L99}}.

\subsection{Stationary values}

\begin{figure}
\begin{tabular}{cc}
 \includegraphics[height=0.29\textheight]{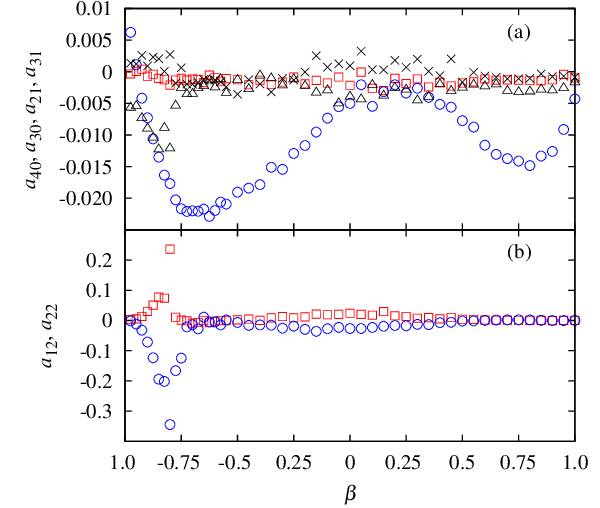}   &
 \includegraphics[height=0.29\textheight]{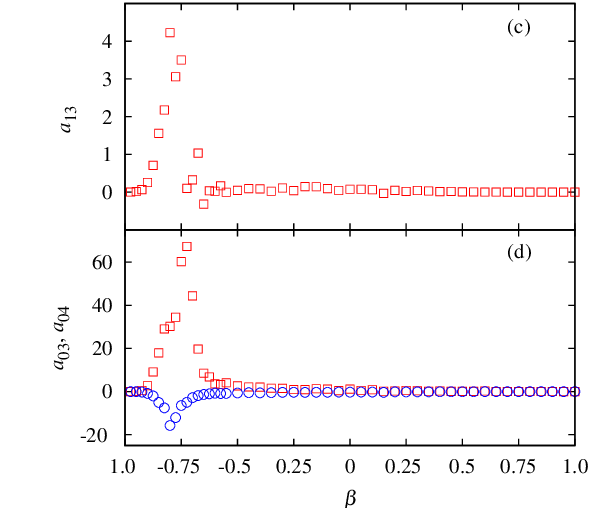}\\
  \caption{DSMC results for the stationary values  {of the sixth- and eighth-order cumulants}, as  functions of roughness  $\beta$, for a granular gas with  {$\alpha=0.9$ and $I=m\sigma^2/10$}. In panel (a),  {crosses, triangles, squares, and  circles stand for $a_{30}$,   $a_{21}$,   $a_{40}$, and   $a_{31}$, respectively}. In panel (b), {circles and squares stand for $a_{12}$ and $a_{22}$, respectively}. In panel (d)  {circles and squares stand for $a_{03}$ and $a_{04}$, respectively}.}
  \label{fig3}
\end{tabular}
\end{figure}

It is interesting to note in Fig.\ \ref{fig2} that  the relaxation time tends to significantly grow in a  roughness region  {close} to the smooth limit $\beta\to-1$.  {One could expect that this signals a maximum in the  {magnitude of the asymptotic} stationary values of the cumulants in the limit $\beta\to-1$, However,  we saw in our previous work \cite{VSK14} that the maxima actually occur  \emph{before}} $\beta=-1$. This picture appears again in the stationary values of the higher-order cumulants analyzed in the present work, as we can see in Fig.\ \ref{fig3}.  {The magnitude of all} these cumulants present a  maximum value in a nearly-smooth region, and approximately  {at} the same value of $\beta$  {as in the case of the fourth-order cumulants. For $\alpha=0.9$ this occurs at} $\beta\simeq -0.775$.

{We can also observe that $|a_{03}|$ and $|a_{04}|$ reach maximum values} that are significantly higher than  the rest of cumulants. Those maxima are actually much higher than that of $|a_{02}|$, which is the largest fourth-order cumulant in   {the \emph{critical}} region.  {This} result is important since it confirms that, at least in this special region of roughness values, the Maxwellian approximation dramatically fails. Actually, the result $|a_{04}|>|a_{03}|>|a_{02}|$  indicates that the  {polynomial expansion}  is  {not convergent} \cite{BP06}. Kinetic theories for granular gases are usually based on perturbative solutions around the Maxwellian and for all cases studied in  {the literature} this approach works  {reasonably well \cite{BP00,HOB00,AP06,AP07,SM09}. To the best of our knowledge, we report for the first time here and in Ref.\ \cite{VSK14} such huge deviations from the Maxwellian for a homogeneous distribution of a granular gas.

\section{Concluding remarks}
Looking back at the results in the present work and in our previous work \cite{VSK14}, we may distinguish between two qualitatively  {different} roughness regions in the  {HCS of a granular gas of inelastic rough hard spheres}. {First, a  region  where the velocity distribution function is \emph{weakly} non-Maxwellian (WNM). In such a case, the distribution can be well represented by  a Sonine polynomial} series expansion around the Maxwellian  {and a linearized Grad approximation works well. We see from Fig.\ \ref{fig3} that with $\alpha=0.9$ this WNM region corresponds to $-1\leq\beta\lesssim -0.9$ and $-0.6\lesssim \beta \leq 1$.}  {In the complementary (and narrower) second region ($-0.9\lesssim\beta\lesssim -0.6$ if $\alpha=0.9$) the velocity distribution is \emph{strongly} non-Maxwellian (SNM), especially in what concerns the distribution of angular velocities. Note that the SNM region was termed ``critical''  above.}
{Figures \ref{fig4}(a) and \ref{fig4}(b) show  representative examples of the velocity distribution function in the WNM and SNM regions, respectively.}

\begin{figure}
\begin{tabular}{cc}
\includegraphics[height=0.21\textheight]{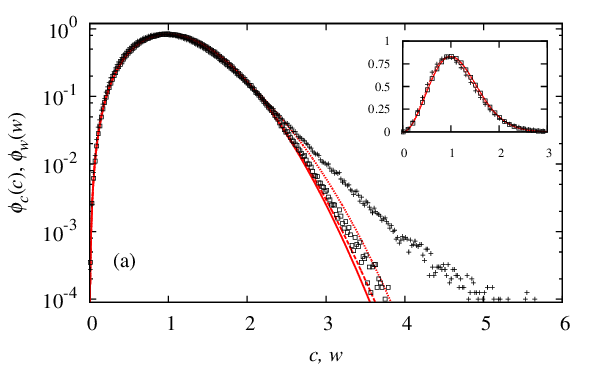}   &
\includegraphics[height=0.21\textheight]{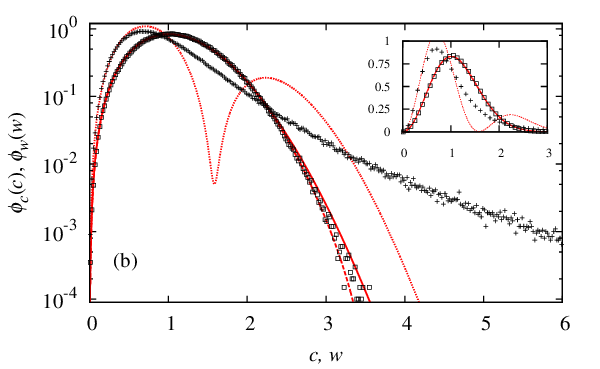}\\
    \caption{DSMC results (in logarithmic scale) for the   marginal distribution functions $\phi_c(c)=4\pi c^2\int d\mathbf{w}\,\phi(\mathbf{c},\mathbf{w})$ ($\square$) and $\phi_w(w)=4\pi w^2\int d\mathbf{c}\,\phi(\mathbf{c},\mathbf{w})$  ($+$) at (a) $\alpha=0.7$, $\beta=0.5$ and (b) $\alpha=0.9$, $\beta=-0.75$.  The solid lines stand for the (common) Maxwellian distribution function, while the dashed and dotted lines stand for the polynomial expansion \eqref{phi} truncated after the fourth-order cumulants in the cases of $\phi_c(c)$ and $\phi_w(w)$, respectively. The insets show the distribution functions in normal scale.}
    \end{tabular}
 \label{fig4}
\end{figure}

{Interestingly enough, as inelasticity increases (for instance, with $\alpha=0.7$), the SNM region widens and its non-Maxwellian features become less remarkable, while those of the WNM region become less negligible \cite{VSK14}. Therefore, an increase of inelasticity tends to smooth out the  borders between the WNM and SNM regions. It is also worthwhile noting that orientational
translational-rotational correlations, which imply $a_{00}^{(1)}\neq 0$  \cite{BPKZ07,VSK14,RM14}, are non-Maxwellian properties that are not especially large in the SNM region \cite{VSK14}.}

{While} previous works already showed that the rough granular gas could be strongly non-Maxwellian \cite{GNB05}, {only} now really huge deviations have been measured and the complete  {spectrum} of roughness values has been explored. Therefore, with our description of the  {fourth-order} cumulants in our previous work \cite{VSK14} and those of  {sixth- and eighth-order in the present work},  we {have  a} global picture of  {what} the  {HCS velocity} distribution function of the granular gas of rough spheres really looks like. Furthermore, the present results make now even more evident that the Maxwellian approximation does not work at all in a  {narrow} interval of roughness values  {(what we called the SNM or critical region)}. This issue cannot  {be} circumvented by just adding more terms in an expansion around the Maxwellian,  as suggested recently \cite{RM14}, since the series is  {likely not} convergent, as we have shown. Outside the  {SNM} region, this kind of Maxwellian-based approach still works. {However,} if we want to eventually attain  {a} complete theoretical description of the  {velocity} distribution function of this type of system, we would need to construct a \emph{novel} approach.

On the other hand, and as we advanced, results in this work also confirm that the  {HCS} hydrodynamic solution, {whether or not close to a Maxwellian,} is always attained. Furthermore, we can confirm that the aging time to hydrodynamics does not increase with increasing inelasticity  {(i.e., decreasing $\alpha$)}. This means that inelastic cooling alone, as long as the homogeneous base state lasts, does not break scale separation (i.e. that hydrodynamic fields variations continue to be slow compared to the microscopic reference time and length scales). This question is of interest since there has been concern about eventual restrictive effects of inelasticity over hydrodynamics. For instance, it sometimes has been used for the HCS an expansion in series of powers of the inelasticity (both for smooth and rough spheres) \cite{SG98,GNB05}.  As a consequence, the hydrodynamic solution is limited to the range of small inelasticities. In other cases, the Chapman-Enskog procedure is inherently limited to the quasi-elastic limit \cite{JR85,GS95}, based on the same concerns of lack of scale separation. However, the results in the present work and in \cite{VSK14} help to clarify that scale separation occurs for the whole range of inelasticities in the HCS.


\begin{theacknowledgments}
F.V.R. and A.S. acknowledge support of the Spanish Government through Grant No.\ FIS2013-42840-P and the Junta de Extremadura through Grant No.\ GR10158, partially financed by FEDER funds. G.M.K. acknowledges support from the Conselho Nacional de Desenvolvimento Cient\'ifico e Tecnol\'ogico (Brazil).
\end{theacknowledgments}

\bibliographystyle{aipproc}
\bibliography{FVR_AS_GK}

\end{document}